\documentclass[aps, a4paper,superscriptaddress,12pt,preprintnumbers,floatfix,nofootinbib,amsmath,amssymb]{revtex4}
\usepackage{amstext,amssymb}
\usepackage{amsmath}
\usepackage{graphicx}
\usepackage{dcolumn}
\usepackage[hyperfootnotes=false]{hyperref}
\usepackage{xspace}
\usepackage{braket}
\usepackage{color}
\usepackage{units}
\usepackage{slashed} 
\usepackage{graphicx}
\usepackage{bm}
\def\L{{\cal L}} 
\def\331{$\mathrm{ SU(3)_c \otimes SU(3)_L \otimes U(1)_X}$}
  
\def\SM{$\mathrm{ SU(3)_c \otimes SU(2)_L \otimes U(1)_Y}$ }

\begin{document}
\title{331 Models and Grand Unification: \\ \vspace{-0.3cm} From Minimal SU(5) to Minimal SU(6)}
\author{\hspace*{-2cm} Frank F. \surname{Deppisch}}
\email{f.deppisch@ucl.ac.uk}
\affiliation{\hspace*{-0.0cm} Department of Physics and Astronomy, University College London, London WC1E 6BT, United Kingdom}
\author{Chandan \surname{Hati}}
\email{chandan@prl.res.in}
\affiliation{\hspace*{-0.0cm}Theoretical Physics Division, Physical Research Laboratory, Ahmedabad 380009, India}
\affiliation{\hspace*{-0.0cm}Discipline of Physics, Indian Institute of Technology Gandhinagar, Ahmedabad 382424, India}
\author{Sudhanwa \hspace*{-0.0cm}\surname{Patra}}
\email{sudha.astro@gmail.com}
\affiliation{Center of Excellence in Theoretical and Mathematical Sciences,
Siksha 'O' Anusandhan University, Bhubaneswar-751030, India}
\author{Utpal \surname{Sarkar}}
\email{utpal@phy.iitkgp.ernet.in}
\affiliation{\hspace*{-0.0cm}Theoretical Physics Division, Physical Research Laboratory, Ahmedabad 380009, India}
\affiliation{\hspace*{-0.0cm}Department of Physics, Indian Institute of Technology Kharagpur, Kharagpur 721302, India}
\author{Jos\`e W.F. \surname{Valle}}
\email{valle@ific.uv.es}
\affiliation{AHEP Group, Institut de Fisica Corpuscular -- C.S.I.C./Universitat de
Valencia, Parc Cientific de Paterna \\
C/~Catedr\`atico Jos\`e Beltr\`an, 2~E-46980 Paterna (Valencia) - Spain}
\begin{abstract}
  We consider the possibility of grand unification of the \331 model
  in an SU(6) gauge unification group. Two possibilities arise.
  Unlike other conventional grand unified theories, in SU(6) one can
  embed the 331 model as a subgroup such that different multiplets
  appear with different multiplicities.  Such a scenario may emerge
  from the flux breaking of the unified group in an E(6) F-theory GUT.
  This provides new ways of achieving gauge coupling unification in
  331 models while providing the radiative origin of neutrino masses.
  Alternatively, a sequential variant of the \331 model can fit
    within a minimal SU(6) grand unification, which in turn can be a
    natural E(6) subgroup.
  This minimal SU(6) embedding does not require any bulk exotics to
  account for the chiral families while allowing for a TeV scale \331
  model with seesaw-type neutrino masses.

\end{abstract}
\maketitle
\newpage
\section{Introduction}

The discovery of the Higgs boson established the existence of spin-0
particles in nature and this opened up the new era in looking for
extensions of the Standard Model (SM) at accelerators. It is now expected
that at higher energies, the SM may be embedded in larger
gauge structures, whose gauge symmetries would have been broken by the
new Higgs scalars. So, we can expect signals of the new gauge bosons,
additional Higgs scalars as well as the extra fermions required to
realize the higher symmetries.
One of the extensions of the SM with the gauge group \331 provides
strong promise of new physics that can be observed at the LHC or the
next generation accelerators
\cite{Singer:1980sw,Valle:1983dk}. Recently there has been a renewed
interest in this model as it can provide novel ways to understand
neutrino masses \cite{Boucenna:2014ela,Boucenna:2014dia}.

The \331 model proposed by Singer, Valle and Schechter (SVS)
\cite{Singer:1980sw} has the special feature that it is not anomaly
free in each generation of fermions, but only when all the three
generations of fermions are included the theory becomes anomaly
free. As a result, different multiplets of the \331 group appear with
different multiplicity and as a result it becomes difficult to unify
the model within usual grand unified theories. For this reason string
completions have been suggested~\cite{Addazi:2016xuh}. In this article
we study how such a theory can be unified in a larger SU(6) gauge
theory that can emerge from a E(6) Grand Unified Theory
(GUT)~\cite{Gursey:1975ki}. We find that the anomaly free
representations of the SVS 331 model can all be embedded in a
combination of anomaly free representations of SU(6), which in turn
can be potentially embedded in the fundamental and adjoint
representations of the group E(6) motivated by F-theory GUTs with
matter and bulk exotics obtained from the flux breaking
mechanism~\cite{1126-6708-2009-01-059,King:2010mq,Callaghan:2011jj,Callaghan:2013kaa}. 

Interestingly,
the SVS 331 model can also be refurbished in an anomaly free multiplet
structure which can be right away embedded in a minimal anomaly free
combination of representations of SU(6) as an E(6) subgroup. We refer
to this new 331 model as the sequential 331 model.  This scheme is
particularly interesting since its embedding in SU(6) does not require
any bulk exotics to account for the chiral families; and in that sense
it provides a truly minimal unification scenario in the same spirit
akin to the minimal SU(5) construction~\cite{Georgi:1974sy}.

The article is organized as follows. In Section~\ref{sec2} we discuss
the basic structure of the SVS \331 model whereas Section~\ref{sec3}
describes the sequential \331 model. In Section~\ref{sec4} we then
analyze the resulting renormalization group running of the gauge
couplings in the SVS model with and without additional octet states,
and discuss necessary conditions for gauge unification. In
Section~\ref{sec5} we then embed the different variants of the \331
model in an SU(6) unification group and demonstrate successful
unification scenarios. Section~\ref{sec6} concerns the experimental
constraints from achieving the correct electroweak mixing angle and
satisfying proton decay limits. We conclude in
Section~\ref{sec:discussion-outlook}.


\section{The SVS \331 Model}{\label{sec2}}

The \331 extension of the SM was originally proposed to
justify the existence of three generations of fermions, as the model
is anomaly free only when three generations are present.
Such a \emph{non-sequential} model, which is generically referred to as the
331 model, breaks down to the SM at some higher energies,
usually expected to be in the TeV range, making the model testable in
the near future. The symmetry breaking: \331 $\to$ \SM allows us to
identify the generators of the 321 model in terms of the generators of
the 331 model. Writing the generators of the $\mathrm{SU(3)_L}$ group
as
\begin{eqnarray}{\label{2.1}} 
  T_3 = \frac{1}{2} I_3 = \begin{pmatrix} \frac{1}{2}&0&0 \\ 0 & - \frac{1}{2} & 0 \\
  0 & 0 & 0\end{pmatrix} \quad &{\rm and }& \quad T_8 = \frac{1}{2 \sqrt{3}} I_8 = \frac{1}{2 \sqrt{3}}
  \begin{pmatrix}1&0&0 \\ 0&1&0 \\ 0&0&-2 \end{pmatrix} \nonumber \\
  {\rm with}~~ I_3 = {\rm diag} [ 1,~-1, ~ 0 ] ~ &{\rm and}&~
  I_8 = {\rm diag} [ 1,~1, ~ -2 ],
\end{eqnarray}
we can readily identify the SM hypercharge and the electric charge as
\begin{equation}{\label{2.2}} 
 Y= \frac{1}{\sqrt{3}} T_8 + X = \frac{1}{6} I_8 + X ~~~ {\rm and} ~~~
 Q = T_3 + Y = \frac{1}{2} I_3 + \frac{1}{6} I_8 + X \,.
\end{equation}
This allows us to write down the fermions and the representations
in which they belong as
\begin{eqnarray}{\label{2.3}} 
 Q_{iL} &=& \begin{pmatrix} u_{iL} \\ d_{iL} \\ D_{iL} \end{pmatrix}  \equiv 
 [3,3,0], \quad
 Q_{3L} = \begin{pmatrix} b_L \\ t_L \\ T_{L} \end{pmatrix}  \equiv 
 [3,{3}^\ast,1/3], \nonumber \\[.15in]
 u_{iR} &\equiv& [3,1,2/3], ~~ d_{iR}  \equiv [3,1,-1/3], ~~ D_{iR} 
 \equiv [3,1,-1/3] \nonumber \\[.15in]
 b_R &\equiv& [3,1,-1/3], ~~ t_{R}  \equiv [3,1,2/3], ~~ T_{R} 
 \equiv [3,1,2/3], \nonumber \\[.15in]
  \psi_{aL} &=& \begin{pmatrix} e_{aL} \\ \nu_{aL} \\ N_{aL} \end{pmatrix} \equiv  [ 1,{3^\ast},-1/3], \quad
 e_{aR} \equiv  [1,1,-1]. 
\end{eqnarray}
The generation index $i=1,2$ corresponds to the first two generations
with the quarks $u_{L,R},d_{L,R},D_{L,R}$ and
$c_{L,R},s_{L,R},S_{L,R}$. For the leptons, the generation index is
$a=1,2,3$.

There are several variants of the model that allow slightly different
choices of fermions as well as their baryon and lepton number
assignments. Here we shall restrict ourselves to the one which
contains only the quarks with electric charge $2/3$ and $1/3$ and no
lepton number ($L$). In this scenario all quarks (usual ones and the
exotic ones) carry baryon number ($B=1/3$) and no lepton number
($L=0$), while all leptons carry lepton number ($L=1$) and no baryon
number ($B=0$). Notice that in Ref. \cite{Boucenna:2014dia} the lepton
number is defined as $L=4/\sqrt{3}~ T_8+ \L$, where $U(1)_{\L}$ is a
global symmetry and a $Z_{2}$ symmetry is introduced to forbid a
coupling like $ \psi_L \psi_L \phi_0$, in connection with neutrino
masses. Since the charge equation given in Eq.~(\ref{2.2}) remains the
same for this assignment, the following discussion regarding
Renormalization Group Equations (RGE) in the SVS model remains valid
for this assignment as well.

For the symmetry breaking and the charged fermion masses, the
following Higgs scalars and their vacuum expectation values (vevs) are
assumed, 
$$ 
\phi_0 \equiv [1,3^\ast, 2/3] ~~~ {\rm and} ~~~ \phi_{1,2}
\equiv [1,3^\ast,-1/3],
$$
\begin{equation}{\label{2.4}} 
 \langle \phi_0 \rangle = \begin{pmatrix} k_0 \\ 0 \\ 0 \end{pmatrix} , ~~~~
 \langle \phi_{1} \rangle = \begin{pmatrix} 0 \\ k_1 \\ n_1\end{pmatrix} , ~~~~
 \langle \phi_2 \rangle = \begin{pmatrix} 0 \\ k_2 \\ n_2\end{pmatrix} \,.
\end{equation}
Here we assume $k_{0,1,2} \sim m_W$ to be of the order of the
electroweak symmetry breaking scale and $n_{1,2} \sim M_{331}$ to be
the \331 symmetry breaking scale. We shall not discuss here the details of
fermion masses and mixing, which can be found in Refs. \cite{Boucenna:2014ela,Boucenna:2014dia}.


\section{The sequential \331 Model}{\label{sec3}}

In this model the fields are assigned in a way such that the anomalies
are cancelled for each generation separately. The multiplet structure
is given by
\begin{eqnarray}{\label{3.1}} 
 Q_{aL} &=& \begin{pmatrix} u_{aL} \\ d_{aL} \\ D_{aL} \end{pmatrix}  \equiv 
 [3,3,0], ~
 u_{aR} \equiv [3,1,2/3], ~ d_{aR}  \equiv [3,1,-1/3], ~ 
 D_{aR} 
 \equiv [3,1,-1/3], \nonumber \\[.15in]
  \psi_{aL} &=& \begin{pmatrix} e^{-}_{aL} \\ \nu_{aL} \\ N^{1}_{aL} \end{pmatrix} \equiv  [ 1,{3^\ast},-1/3], ~
  \xi_{aL} = \begin{pmatrix} E^{-}_{aL} \\ N^{2}_{aL} \\ N^{3}_{aL} \end{pmatrix} \equiv  [ 1,{3^\ast},-1/3], ~
    \chi_{aL} = \begin{pmatrix} N^{4}_{aL} \\ E^{+}_{aL} \\ e^{+}_{aL} \end{pmatrix} \equiv  [ 1,{3^\ast},2/3].\nonumber\\
\end{eqnarray}

It is straightforward to check that each family is anomaly free. In
order to drive symmetry breaking and generate the charged fermion
masses, we assume a Higgs sector and vevs similar to the SVS 331 model
\footnote{A model with similar fermion content and with
  $k_{1}=n_{2}=0$ in the scalar sector was discussed in
  Ref. \cite{Sanchez:2001ua} using the trinification group 
$\mathrm{SU(3)_c \times SU(3)_L \times SU(3)_R}$.}.
The Yukawa Lagrangian for the quark sector can be written as
\begin{eqnarray}{\label{3.2}} 
\mathcal{L}_{\rm{quarks}}=y_{u_a} \overline{Q_{aL}} u_{aR}\phi_{0}^{\ast}+y_{d_a}^{i} \overline{Q_{aL}} d_{aR}\phi_{i}^{\ast}+y_{D_a}^{i} \overline{Q_{aL}} D_{aR}\phi_{i}^{\ast} + {\rm{h.c.}} \quad,
\end{eqnarray}
with $i=1,2$ and where we neglect any flavour mixing. After the chain of spontaneous symmetry breaking the
up-type quarks obtain a mass term 
\[
m_{u_a}=y_{u_a} k_{0},
\]
while the down--type and vectorlike down--type quarks form a mass matrix
in the $(d,D)$ basis given by
\begin{eqnarray}{\label{3.3}}
m_{dD}^{a}=\begin{pmatrix} y_{d_a}^{1} k_1 +y_{d_a}^{2} k_2 & y_{D_a}^{1} k_1 +y_{D_a}^{2} k_2\\
 y_{d_a}^{1} n_1 +y_{d_a}^{2} n_2 & y_{D_a}^{1} n_1 +y_{D_a}^{2} n_2\end{pmatrix} .
 \end{eqnarray}
 Note that in the cases
 $y_{d_a}^{1}=y_{d_a}^{2}=y_{D_a}^{1}=y_{D_a}^{2}\equiv y_d$; or
 $k_1=k_2=k$ and $n_1=n_2=n$ the determinant of the above Yukawa
 matrix vanishes giving $m_{d_a}=0$ and
 $m_{D_a}=y_{d_a}^{1} k_1 +y_{d_a}^{2} k_2 +y_{D_a}^{1} n_1
 +y_{D_a}^{2} n_2$. However, in the absence of any symmetries forcing
 the above conditions, the down quarks obtain mass as a result of the
 mixing with the vector--like quarks. One can determine it
 perturbatively by expanding the Yukawa contributions in terms of
 $k_i/n_i \ll 1$ so as to obtain
$$m_{d_a}=\left(y_{d_a}^{1} k_1 +y_{d_a}^{2} k_2 \right) -\left( y_{D_a}^{1} k_1 +y_{D_a}^{2} k_2 \right) \frac{y_{d_a}^{1} n_1 +y_{d_a}^{2} n_2}{y_{D_a}^{1} n_1 +y_{D_a}^{2} n_2}+\cdots ,$$
 $$m_{D_a}=\left(y_{d_a}^{1} k_1 +y_{d_a}^{2} k_2 \right)+\left(y_{D_a}^{1} n_1 +y_{D_a}^{2} n_2\right)-\left| M_{dD}^{a} \right|/ \left(y_{D_a}^{1} n_1 +y_{D_a}^{2} n_2\right) +\cdots ,$$
  where 
 $$\left| M_{dD}^{a} \right|=\left(y_{d_a}^{1} k_1 +y_{d_a}^{2} k_2 \right)\left(y_{D_a}^{1} n_1 +y_{D_a}^{2} n_2\right)-\left( y_{D_a}^{1} k_1 +y_{D_a}^{2} k_2 \right)\left( y_{d_a}^{1} n_1 +y_{d_a}^{2} n_2 \right).$$ 

 This structure can be used to account for the SM quark masses and CKM
 mixing, as well as the heavier vector--like quark mass limits from
 the LHC.

 Turning now to the lepton sector, the relevant Yukawa terms are given by
\begin{eqnarray}{\label{3.4}} 
\mathcal{L}_{\rm{leptons}}=\epsilon_{\alpha\beta\gamma}\left[ \psi_{\alpha L}^{T} C^{-1}\left(y_{1}  \xi_{\beta L}\phi_{0\gamma}+y_{2}^{i}\chi_{\beta L}\phi_{i \gamma}\right)+\xi_{\alpha L}^{T} C^{-1} y_{3}^{i} \chi_{\beta L}\phi_{i \gamma} \right] + {\rm{h.c.}} \quad,
\end{eqnarray}
where $\alpha, \beta, \gamma$ are the $\mathrm{SU(3)_{L}}$ tensor
indices ensuring antisymmetric Dirac mass terms, $C$ is the charge
conjugation matrix, and $i=1,2$. After the symmetry breaking, these
Yukawa terms give rise to the mass matrices for charged and neutral
leptons. In the basis $(e,E)$ the mass matrix is given by
\begin{eqnarray}{\label{3.5}}
m_{eE}=\begin{pmatrix} -\left(y_{2}^{1} k_1 +y_{2}^{2} k_2\right) & \left(y_{2}^{1} n_1 +y_{2}^{2} n_2\right)\\
 -\left(y_{3}^{1} k_1 +y_{3}^{2} k_2 \right)& \left( y_{3}^{1} n_1 +y_{3}^{2} n_2 \right)\end{pmatrix} ,
 \end{eqnarray}
with the eigenvalues given by  
$$m_{e}=-\left(y_{2}^{1} k_1 +y_{2}^{2} k_2\right) +\left(y_{3}^{1} k_1 +y_{3}^{2} k_2 \right) \frac{y_{2}^{1} n_1 +y_{2}^{2} n_2}{y_{3}^{1} n_1 +y_{3}^{2} n_2}+\cdots ,$$
$$m_{E}=\left( y_{3}^{1} n_1 +y_{3}^{2} n_2 \right)-\left(y_{2}^{1} k_1 +y_{2}^{2} k_2\right)-\left| M_{eE}^{a} \right|/ \left( y_{3}^{1} n_1 +y_{3}^{2} n_2 \right) +\cdots ,$$
where
$$\left| M^{a}_{eE}\right|=\left(y_{2}^{1} n_1 +y_{2}^{2}
  n_2\right)\left(y_{3}^{1} k_1 +y_{3}^{2} k_2 \right)-\left(y_{2}^{1}
  k_1 +y_{2}^{2} k_2\right)\left( y_{3}^{1} n_1 +y_{3}^{2} n_2
\right).$$ 

For the case of neutral leptons the mass matrix can be written as:
\begin{eqnarray}{\label{3.6}}
m_{\nu N}=\begin{pmatrix} 
0 & 0 & y_1 k_0  & 0 & -\left(y_{2}^{1} n_1 +y_{2}^{2} n_2\right)\\
 0 & 0 & 0 & -y_1 k_0 & \left(y_{2}^{1} k_1 +y_{2}^{2} k_2\right) \\
  y_1 k_0 & 0 & 0 & 0 & \left(y_{3}^{1} k_1 +y_{3}^{2} k_2 \right) \\
 0 & -y_1 k_0 & 0 & 0 & -\left(y_{3}^{1} n_1 +y_{3}^{2} n_2 \right) \\
-\left(y_{2}^{1} n_1 +y_{2}^{2} n_2\right) & \left(y_{2}^{1} k_1 +y_{2}^{2} k_2\right) & \left(y_{3}^{1} k_1 +y_{3}^{2} k_2 \right)  &-\left(y_{3}^{1} n_1 +y_{3}^{2} n_2 \right) & 0
  \end{pmatrix} ,\nonumber\\
 \end{eqnarray}
 in the basis $(\nu, N_1, N_3, N_2, N_4)$, where $N_1, N_3$ are
 SU(2)$_{\rm L}$ isosinglets and $\nu, N_2, N_4$ are components of doublets. Next,
 we rotate the above mass matrix by an orthogonal transformation
 $m_{\nu N}^{\prime}=R^{T} m_{\nu N} R$, where
\begin{eqnarray}{\label{3.6.1}}
R=\begin{pmatrix} 
0 &- \frac{1}{2} & \frac{1}{2} & \frac{1}{2} & \frac{1}{2}\\
 - \frac{1}{\sqrt{2}} & \frac{1}{2} &\frac{1}{2} & 0 & 0 \\
\frac{1}{\sqrt{2}} & \frac{1}{2} & \frac{1}{2} & 0 & 0 \\
0 & \frac{1}{2} & -\frac{1}{2}& \frac{1}{2} & \frac{1}{2} \\
 0 & 0 & 0 & \frac{1}{\sqrt{2}} & -\frac{1}{\sqrt{2}}
  \end{pmatrix} ,\nonumber\\
 \end{eqnarray}
This yields the rotated mass matrix $m^{\prime}_{\nu N}$ given by
\begin{eqnarray}{\label{3.7}}
m^{\prime}_{\nu N}=\begin{pmatrix} 
0 & 0 & 0 & \frac{u}{\sqrt{2}}-\frac{x-z} {2}& \frac{u}{\sqrt{2}}+\frac{x-z} {2}\\
 0 & -u & 0 & \frac{(X-Z)+(x+z)} {2\sqrt{2}} & -\frac{(X-Z)+(x+z)} {2\sqrt{2}} \\
 0 & 0 & u & -\frac{(X-Z)-(x+z)} {2\sqrt{2}} & \frac{(X-Z)-(x+z)} {2\sqrt{2}} \\
\frac{u}{\sqrt{2}}-\frac{x-z} {2} & \frac{(X-Z)+(x+z)} {2\sqrt{2}} & -\frac{(X-Z)-(x+z)} {2\sqrt{2}} & -\frac{X+Z}{\sqrt{2}} & 0 \\
\frac{u}{\sqrt{2}}+\frac{x-z} {2} & -\frac{(X-Z)+(x+z)} {2\sqrt{2}}  & \frac{(X-Z)-(x+z)} {2\sqrt{2}} & 0 & \frac{X+Z}{\sqrt{2}}
 \end{pmatrix} ,\nonumber\\
 \end{eqnarray}
 where
 $$u=y_1 k_0,~ X= \left(y_{2}^{1} n_1 +y_{2}^{2} n_2\right),
 ~ x= \left(y_{2}^{1} k_1 +y_{2}^{2} k_2\right), ~
 z=\left(y_{3}^{1} k_1 +y_{3}^{2} k_2 \right), ~ Z=\left(y_{3}^{1}
   n_1 +y_{3}^{2} n_2 \right). $$ 
Now we recall that
 $k_{0,1,2} \sim m_W$ is of the order of the electroweak symmetry
 breaking scale, while and $n_{1,2} \sim M_{331}$ is of the order of
 the \331 symmetry breaking scale, and hence one expects that
 $X,Z\gg u, x, z$. If we further assume $X+Z\gg X-Z$, then we can
 identify the 44 and 55 entries as the heaviest in the mass matrix
 given in Eq. (\ref{3.7}) and these rotated isodoublet states form a
 pair of heavy quasi Dirac neutrinos with mass of the order of the
 \331 symmetry breaking scale. We can now readily use perturbation
 theory to obtain the masses for the three remaining lighter
 states. Up to second order in perturbation theory we obtain two Dirac
 states with mass of the order of the electroweak symmetry breaking
 scale $\pm u=\pm y_1 k_0$ and a light seesaw Majorana neutrino with
 mass $2u(z-x)/(X+Z)$. With this we see that the model has enough
 flexibility to account for the observed pattern of fermion masses.
 It is not our purpose here to present a detailed study of the
 structure of the fermion mass spectrum, but only to check its
 consistency in broad terms.


 \section{Renormalization Group Equations and 
Gauge Coupling Unification} {\label{sec4}}

In this section we study the SVS model RGEs to explore if unification of the three gauge
couplings~\cite{PhysRevLett.33.451} can be obtained in the \331 theory
at a certain scale $M_U$, without any presumptions about the nature
of the underlying group of grand
unification~\cite{Boucenna:2014dia}. Using the RGEs we express the
hypercharge (and X) normalization and the unification scale as a
function of \331 breaking scale. Next we study the allowed range of
\331 breaking scale such that one can obtain a guaranteed unification
of the gauge couplings. First we discuss the SVS model discussed in
section \ref{sec2}. Then, we study the impact of adding three
generations of leptonic octet representations $[1,8,0]$ that can give
gauge coupling unification for a TeV scale \331 breaking while driving
an interesting radiative model for neutrino mass
generation~\cite{Boucenna:2014dia}.

The evolution for running coupling constants at one loop level is
governed by the RGEs
\begin{equation}{\label{4.1}}
\mu\,\frac{\partial g_{i}}{\partial \mu}=\frac{b_i}{16 \pi^2} g^{3}_{i},
\end{equation}
which can be written in the form
\begin{equation}{\label{4.2}}
\frac{1}{\alpha_{i}(\mu_{2})}=\frac{1}{\alpha_{i}(\mu_{1})}-\frac{b_{i}}{2\pi} \ln \left( \frac{\mu_2}{\mu_1}\right),
\end{equation}
where $\alpha_{i}=g_{i}^{2}/4\pi$ is the fine structure constant for
$i$--th gauge group, $\mu_1, \mu_2$ are the energy scales with
$\mu_2 > \mu_1$. The beta-coefficients $b_i$ determining the evolution
of gauge couplings at one-loop order are given by
\begin{eqnarray}{\label{4.3}}
	&&b_i= - \frac{11}{3} \mathcal{C}_{2}(G) 
				 + \frac{2}{3} \,\sum_{R_f} T(R_f) \prod_{j \neq i} d_j(R_f) 
  + \frac{1}{3} \sum_{R_s} T(R_s) \prod_{j \neq i} d_j(R_s).
\label{oneloop_bi}
\end{eqnarray}
Here, $\mathcal{C}_2(G)$ is the quadratic Casimir operator for the
gauge bosons in their adjoint representation,
\begin{equation}{\label{4.4}}
	\mathcal{C}_2(G) \equiv \left\{
	\begin{matrix}
		N & \text{if } SU(N), \\
    0 & \text{if }  U(1).
	\end{matrix}\right.
\end{equation}
On the other hand, $T(R_f)$ and $T(R_s)$ are the Dynkin indices of the
irreducible representation $R_{f,s}$ for a given fermion and scalar,
respectively, 
\begin{equation}{\label{4.5}}
	T(R_{f,s}) \equiv \left\{
	\begin{matrix}
		1/2 & \text{if } R_{f,s} \text{ is fundamental}, \\
    N   & \text{if } R_{f,s} \text{ is adjoint}, \\
		0   & \text{if } R_{f,s} \text{ is singlet},
	\end{matrix}\right.
\end{equation}
and $d(R_{f,s})$ is the dimension of a given representation $R_{f,s}$
under all gauge groups except the $i$-th~gauge group under
consideration. An additional factor of $1/2$ is multiplied in the case
of a real Higgs representation.

The electromagnetic charge operator is given by
\begin{equation}{\label{4.6}}
 Q = T_3 + Y =  T_3 +\frac{1}{\sqrt{3}} T_8 + X,
\end{equation}
where the generators (Gell-Mann matrices) are normalized as
$\text{Tr} (T_{i}T_{j})=\frac{1}{2}\delta_{ij}$. We define the
normalized hypercharge operator $Y_{N}$ and $X_{N}$ as
\begin{equation}{\label{4.7}}
Y=n_{Y} Y_{N},\quad X=n_{X}X_{N},
\end{equation}
such that we have
\begin{equation}{\label{4.8}}
n_{Y}^{2}=\frac{1}{3}+n_{X}^{2},
\end{equation}
and the normalized couplings are related by
\begin{equation}{\label{4.9}}
n_{Y}^{2} {\left(\alpha^{N}_{Y}\right)}^{-1}=\frac{1}{3}\alpha_{3L}^{-1}+\left(n_{Y}^{2}-\frac{1}{3}\right){\left(\alpha^{N}_{X}\right)}^{-1},
\end{equation}
where
\begin{equation}{\label{4.10}}
\alpha^{N}_{Y} =n_{Y}^{2} \alpha_{Y}, \quad \alpha^{N}_{X}=\left(n_{Y}^{2}-\frac{1}{3}\right)\alpha_{X}, \quad \alpha_{3L}=\alpha_{2L}.
\end{equation}
Now using Eqs. (\ref{4.1}, \ref{4.9}, \ref{4.10}) we obtain
\begin{eqnarray}{\label{4.11}}
\alpha^{-1}_{U} &=& \frac{1}{n_{Y}^{2}-\frac{1}{3}} 
\left\{
	\alpha_{\text{em}}^{-1}(M_{Z})\cos^{2}\theta_{w} (M_{Z})
	-\frac{1}{3}\alpha^{-1}_{2L}(M_Z)
	-\frac{b^{\text{UN}}_{Y}-\frac{1}{3}b_{2L}}{2\pi}
	 \ln\left(\frac{M_X}{M_Z}\right)
	-\frac{b^{\text{UN}}_{X}}{2\pi}
	 \ln\left(\frac{M_U}{M_X}\right)
\right\},\nonumber\\
\alpha^{-1}_{U}&=&\alpha^{-1}_{2L}(M_Z)-\frac{b_{2L}}{2\pi}\ln\left(\frac{M_X}{M_Z}\right)-\frac{b_{3L}}{2\pi}\ln\left(\frac{M_U}{M_X}\right),\nonumber\\
\alpha^{-1}_{U}&=&\alpha^{-1}_{3C}(M_Z)-\frac{b_{3C}}{2\pi}\ln\left(\frac{M_X}{M_Z}\right)-\frac{b^{X}_{3C}}{2\pi}\ln\left(\frac{M_U}{M_X}\right).
\end{eqnarray}
Here, the SM running is described by the the SU(3)$_C$ coefficient $b_{3C}$, the SU(2)$_L$ coefficient $b_{2L}$ and the U(1)$_Y$ unnormalized coefficient $b^{\text{UN}}_{Y}$. Likewise, in the unbroken \331 phase, the gauge running coefficients for the SU(3)$_C$, SU(3)$_L$ and unnormalized U(1)$_X$ components are $b^{X}_{3C}$, $b_{3L}$ and $b^{\text{UN}}_{X}$, respectively. The scale $M_{Z}$ corresponds to the $Z$
boson-pole, the 331 symmetry breaking scale is denoted by $M_{X}$ and
$M_{U}$ is the scale of unification for the normalized gauge
couplings. From the above set of equations the unification scale $M_U$
can be obtained as a function of $M_{X}$,
\begin{equation}{\label{4.12}}
M_U=M_{X}\left(\frac{M_X}{M_Z}\right)^{-\frac{b_{3C}-b_{2L}}{b^{X}_{3C}-b_{3L}}} \exp \left[2\pi \frac{\alpha^{-1}_{3C}(M_Z)-\alpha^{-1}_{2L}(M_Z)}{b^{X}_{3C}-b_{3L}}\right].
\end{equation}
Similarly, $n_{Y}^{2}$ can be expressed as a function of $M_{X}$,
\begin{eqnarray}{\label{4.13}}
n_{Y}^{2}&=&\frac{1}{3}+\left[\alpha_{\text{em}}^{-1}(M_{Z})\cos^{2}\theta_{w} (M_{Z})-\frac{1}{3}\alpha^{-1}_{2L}(M_Z)-\frac{b^{\text{UN}}_{Y}-\frac{1}{3}b_{2L}}{2\pi}\ln\left(\frac{M_X}{M_Z}\right)\right.\nonumber\\
&+&\left. b^{\text{UN}}_{X}\left\{ \frac{1}{2\pi}\frac{b_{3C}-b_{2L}}{b^{X}_{3C}-b_{3L}}\ln\left(\frac{M_X}{M_Z}\right)- \frac{\alpha^{-1}_{3C}(M_Z)-\alpha^{-1}_{2L}(M_Z)}{b^{X}_{3C}-b_{3L}}\right\}\right]\nonumber\\
&\times& \left[\alpha^{-1}_{2L}(M_Z)-\frac{b_{2L}}{2\pi}\ln\left(\frac{M_X}{M_Z}\right)+b_{3L}\left\{\frac{1}{2\pi}\frac{b_{3C}-b_{2L}}{b^{X}_{3C}-b_{3L}}\ln\left(\frac{M_X}{M_Z}\right)-\frac{\alpha^{-1}_{3C}(M_Z)-\alpha^{-1}_{2L}(M_Z)}{b^{X}_{3C}-b_{3L}}\right\} \right]^{-1}.\nonumber\\
\end{eqnarray}
The above two relations are valid provided $b^{X}_{3C}\neq b_{3L}$ and
$(b^{X}_{3C}-b_{3L})\neq (b_{3C}-b_{2L})$, which are satisfied in the
cases that we shall discuss below. Furthermore, we take $M_{X}\leq
M_{U}\leq 10^{17}$~GeV and assume that 331 is the only gauge group (in
other words $M_{X}$ is the only intermediate scale) between $M_{Z}$
and the unification scale $M_U$.

\subsection{The minimal SVS  Model}

The first case of interest is the minimal scenario described in
section \ref{sec2}. The relevant gauge quantum numbers are given in
Eqs. (\ref{2.3},\ref{2.4}). The Higgs sector involves three
$\mathrm{SU(3)_{L}}$ triplets, namely the minimal set necessary for
adequate symmetry breaking and generation of fermion masses. First we
notice that the model described in Ref. \cite{Boucenna:2014dia} has
the same RGE evolution, since the extra gauge singlets added to the
fermion spectrum to generate neutrino masses do not enter the
RGEs. For the SM the one-loop beta-coefficients are given by
$b_{2L} = -19/6$, $b^{\text{UN}}_{Y} = 41/6$, $b_{3C} = -7$, while in
the $SU(3)_c \times SU(3)_L \times U(1)_X$ phase they are given by
$b_{3L} = -13/2$, $b^{\text{UN}}_{X} = 26/3$, $b^{X}_{3C} = -5$.
\begin{figure}[t!]
\includegraphics[width=0.49\linewidth]{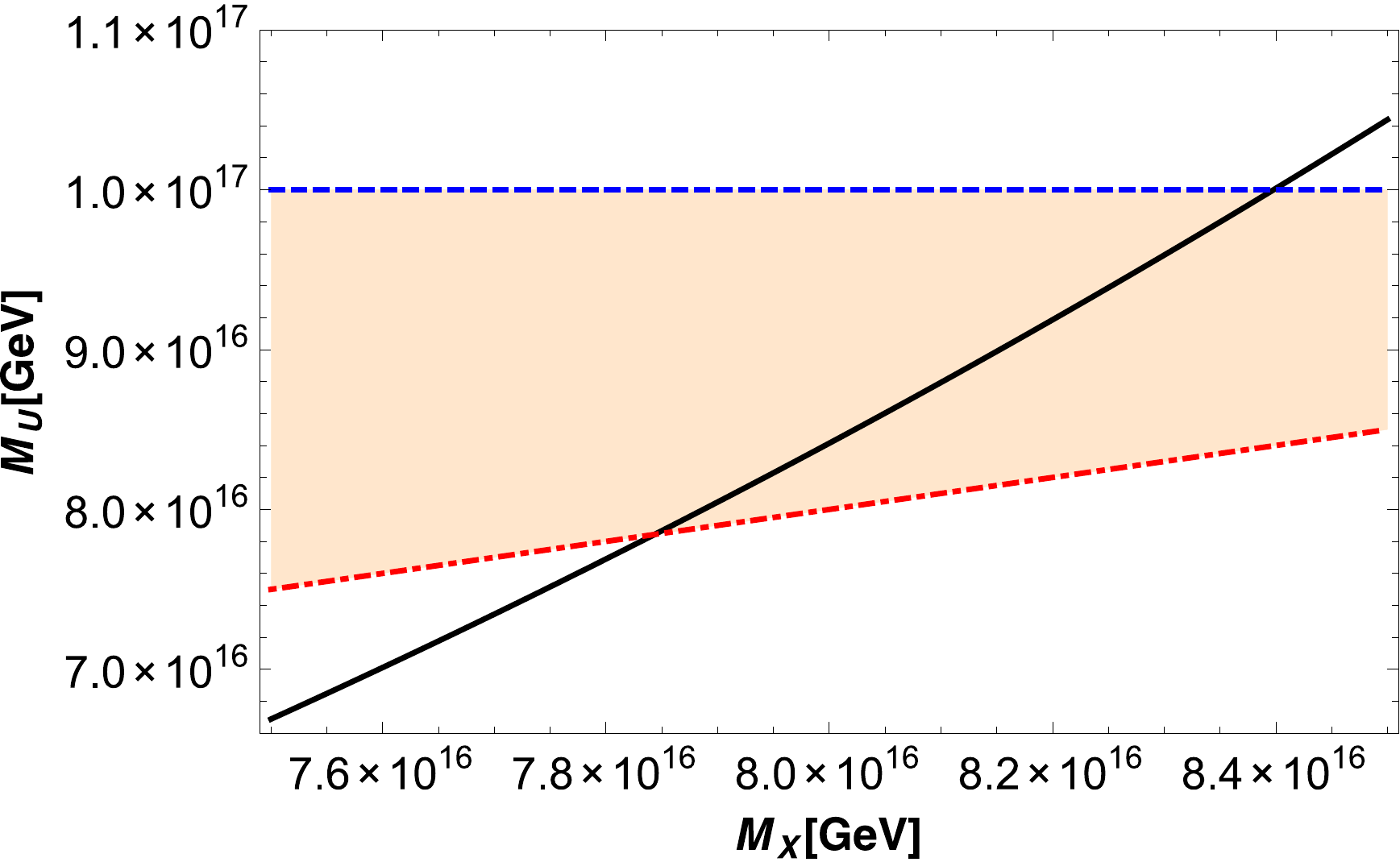}
\includegraphics[width=0.47\linewidth]{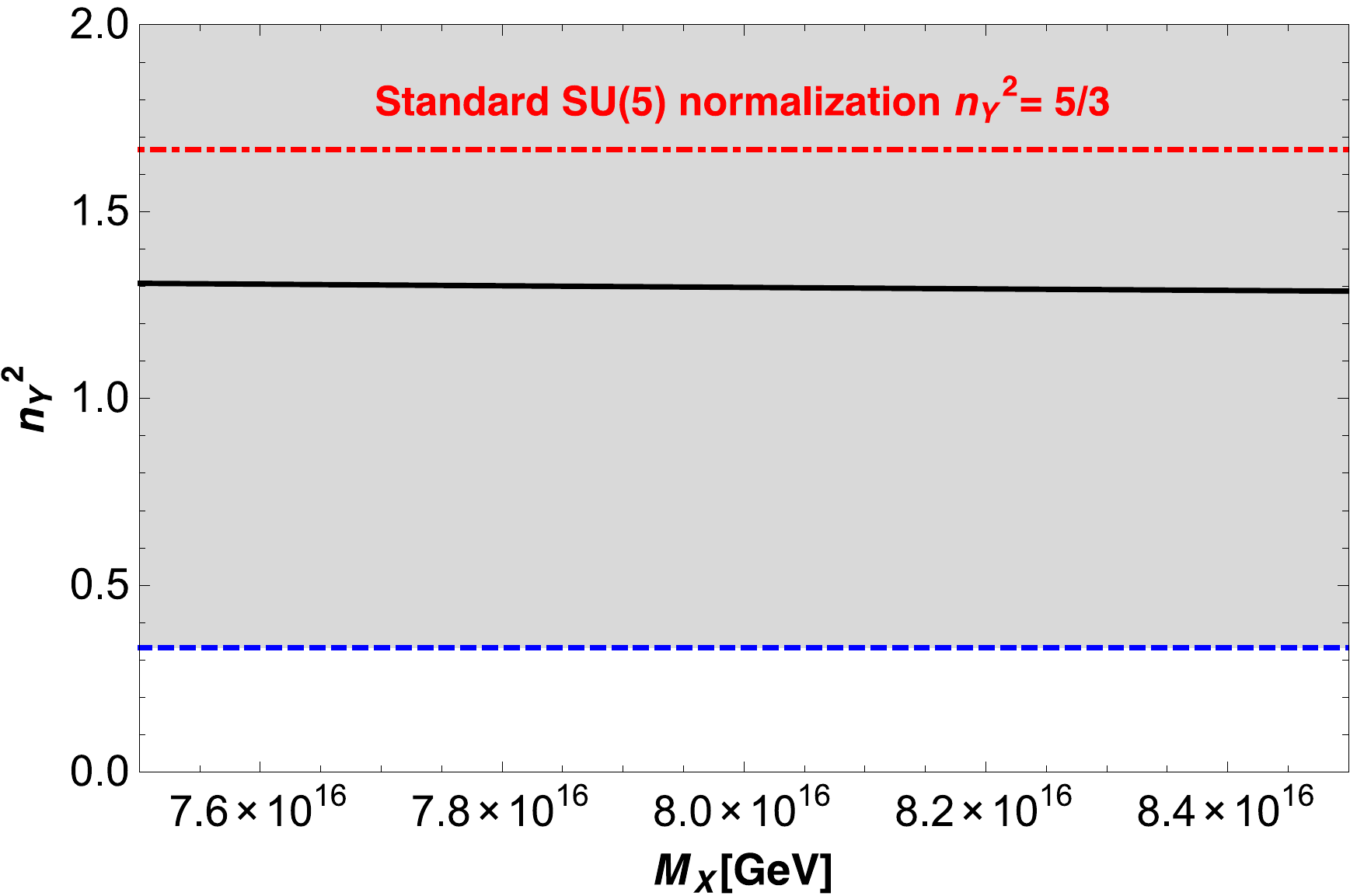}
\caption{(Left) Allowed range for \331 symmetry breaking scale
  $M_{X}$ for guaranteed unification. The solid line represents
  $M_{U}$ as a function of $M_{X}$. The dashed and dot-dashed lines
  correspond to $M_{U}=10^{17}$~GeV and $M_{U}=M_{X}$
  respectively. (Right) The hypercharge normalization factor
  $n_{Y}^{2}$ as a function of the 331 symmetry breaking scale
  $M_{X}$. The shaded region represents the allowed region $n_{Y}^{2}\geq
  \frac{1}{3}$ with the dashed line corresponding to the lower limit
  $n_{Y}^{2}= \frac{1}{3}$. The solid line gives $n_{Y}^{2}$ as a
  function of $M_X$ and the red dot-dashed line shows the standard $SU(5)$ normalization $n_Y^2=
  \frac{5}{3}$.}
\label{fig1}
\end{figure}

In Fig.~\ref{fig1} (left) we plot the allowed range for $M_X$. The intersection of the line corresponding to $M_U$
evaluated as a function of $M_{X}$ in Eq. (\ref{4.12}) with the lines
for $M_{U}=M_{X}$ and  $M_{U}=10^{17}$~GeV gives the lower and upper
bound on $M_{X}$ respectively such that there is a guaranteed
unification. In this scenario, the scale $M_X$ of \331 breaking is therefore always high and very close to the unification scale $M_U$. 

Next, in Fig. \ref{fig1} (right) we plot the hypercharge normalization
factor $n_{Y}^{2}$ as a function of \331 symmetry breaking scale
$M_{X}$. The dashed horizontal line represents the lower limit
$n_{Y}^{2}= \frac{1}{3}$ of the allowed value for $n_{Y}^{2}$. As can
be seen from the figure, for the allowed $M_X$ range from the
condition $M_{X}\leq M_{U}\leq 10^{17}$~GeV the hypercharge
normalization $n_{Y}^{2}$ is almost constant $\approx 1.3$ and well above the allowed lower limit.

\begin{figure}[t!]
\includegraphics[width=\linewidth]{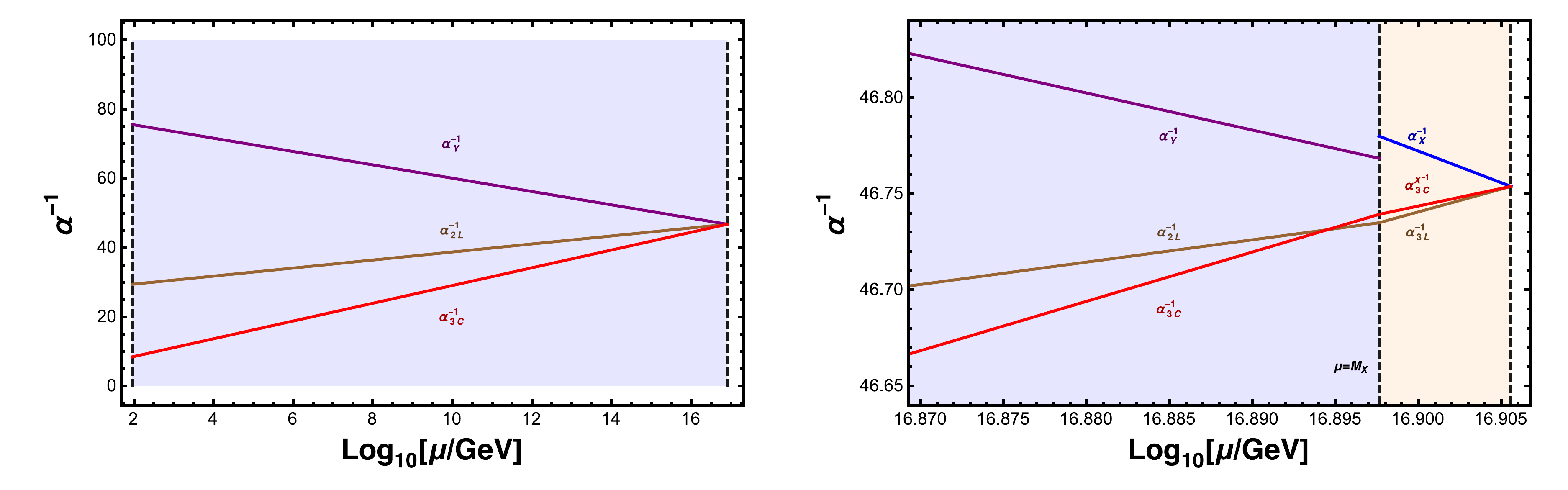}
\caption{Gauge coupling running in the SVS Model with 331 symmetry
  breaking scale $M_{X}=7.9\times 10^{16}$~GeV, demonstrating
  successful gauge unification at the scale
  $M_{U} = 8.05 \times10^{16}$~GeV with $n_{Y}^{2}=1.3$ . The right
  plot shows the magnified view of gauge coupling running around
  $M_{X}$.}
\label{fig3}
\end{figure}

Finally, in Fig.~\ref{fig3} we give an example of gauge coupling
running with respect to the 331 symmetry breaking scale
$M_{X}=7.9\times 10^{16}$~GeV. It demonstrates that successful gauge
unification at the scale $M_{U} = 8.05 \times 10^{16}$~GeV with
$n_{Y}^{2}=1.3$ can be achieved, albeit this requires a very high scale of \331 breaking very near to the unification scale.  

\subsection{The SVS Model with fermionic octets}

In this model, in addition to the field content of model I, we include
three generations of fermion octets $\Omega$ with the assignments
under the \331 group given by
\begin{equation}{\label{4.15}}
 \Omega\equiv [1, 8^{\ast}, 0]. 
\end{equation}

The Higgs sector involves the same three $\mathrm{SU(3)_{L}}$
triplets as before. Although this model has the same content as the one
considered in Ref.~\cite{Boucenna:2014dia}, here we take a completely
different approach to unification. Indeed, we do not consider the
usual SU(5) normalization for the hypercharge and the octet mass scale
is the same as the 331 symmetry breaking scale.  In this model, the
neutrinos are massless at tree level, however at one-loop level the
exchange of gauge bosons give rise to dimension-nine operator which
generates neutrino masses after 331 symmetry
breaking~\cite{Boucenna:2014dia}.  For the SM the one-loop
beta-coefficients remain the same as Model I, while in the \331 phase
they are given by $b_{3L} = -1/2$, $b^{\text{UN}}_{X} = 26/3$,
$b^{X}_{3C} = -5$.

\begin{figure}[t!]
\includegraphics[width=0.49\linewidth]{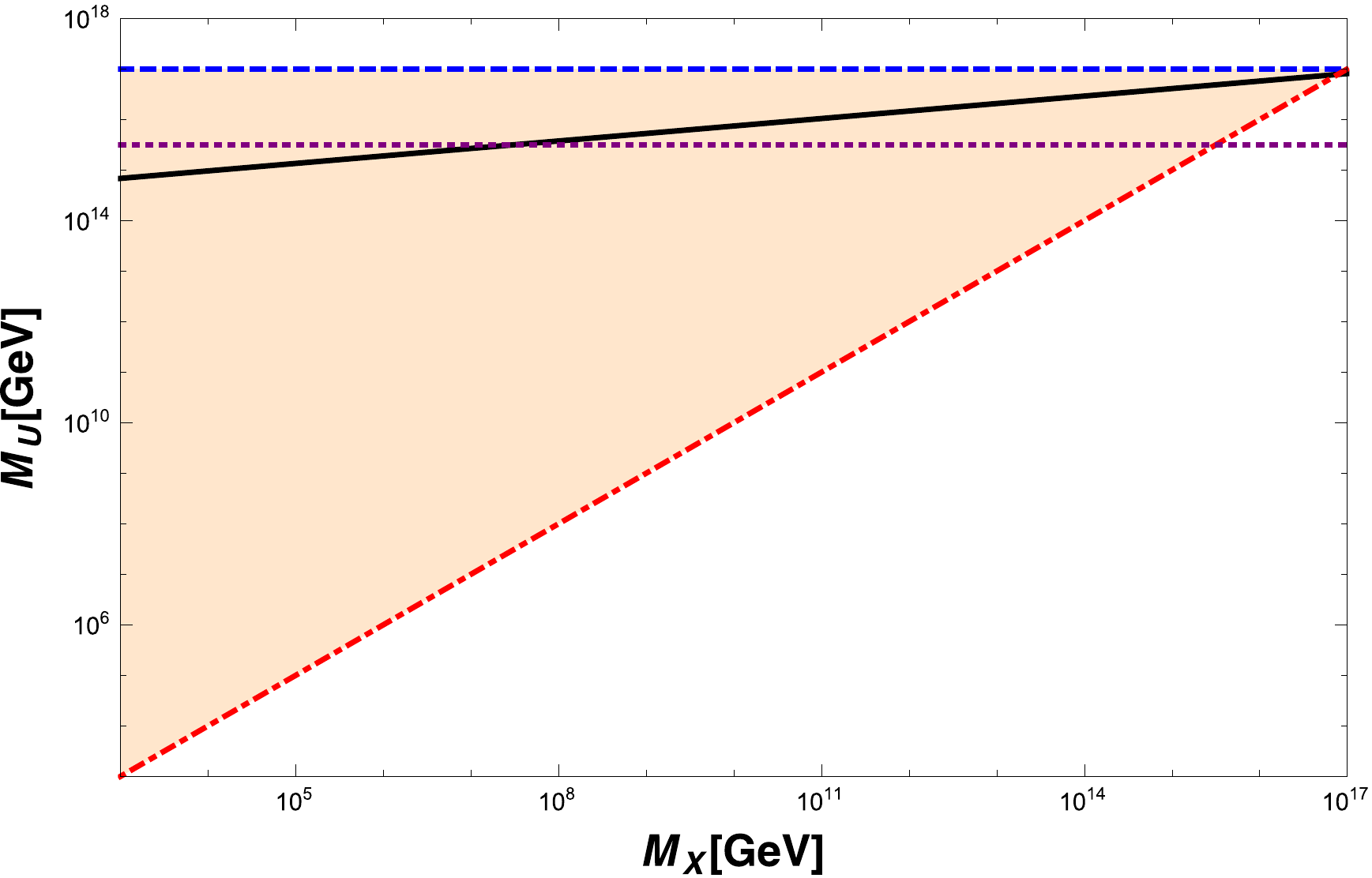}
\includegraphics[width=0.49\linewidth]{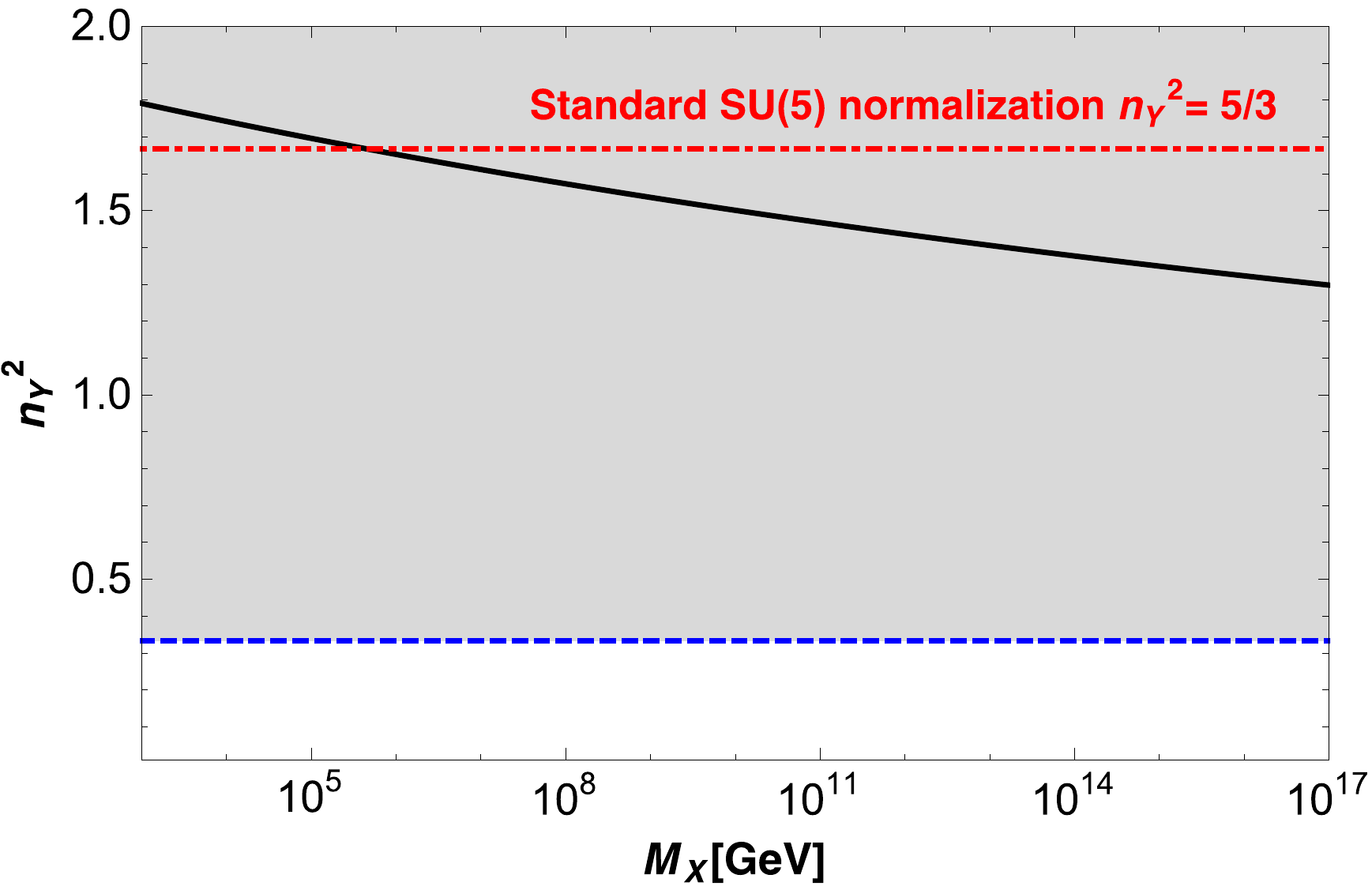}
\caption{Same as Fig.~\ref{fig1}, but for the \331 SVS model with three additional fermionic octets. The purple dotted line corresponds to the lower limit of $M_U$ allowed by the current experimental limits on the lifetime of the proton decay.}
\label{fig7}
\end{figure}

In Fig. \ref{fig7} (left) we plot the allowed range for $M_X$ for
which unification is guaranteed at a scale
$M_{X}\leq M_{U}\leq 10^{17}$GeV. Interestingly, in this model we find
that for a \331 symmetry breaking scale $M_{X}$ as low as TeV it is
possible to achieve unification. Note that in contrast to
Ref. \cite{Boucenna:2014dia}, here we do not assume another
intermediate scale corresponding to the fermion octet mass scale in
addition to $M_{X}$. Formally, unification can thus be achieved for
any scale $M_X$ between $M_Z$ and $M_U$, however,
$M_{U}\lesssim 10^{15.5}$~GeV is disfavored by the current
experimental limits on the lifetime of the proton decay
\cite{Agashe:2014kda}. This consequently puts a lower limit of
  the order of $M_X \gtrsim 10^5$~GeV on the \331 breaking scale,
  although we should emphasize that we here do not specify the GUT
  group and thus cannot predict the proton decay rate accurately.

In Fig. \ref{fig7} (right) we plot the hypercharge normalization
factor $n_{Y}^{2}$ as a function of \331 symmetry breaking scale
$M_{X}$. In this case as well, for the allowed $M_X$ range from the
condition $M_{X}\leq M_{U}\leq 10^{17}$~GeV the hypercharge
normalization $n_{Y}^{2}$ is well above its allowed lower limit. 

\begin{figure}[t!]
\includegraphics[width=0.7\linewidth]{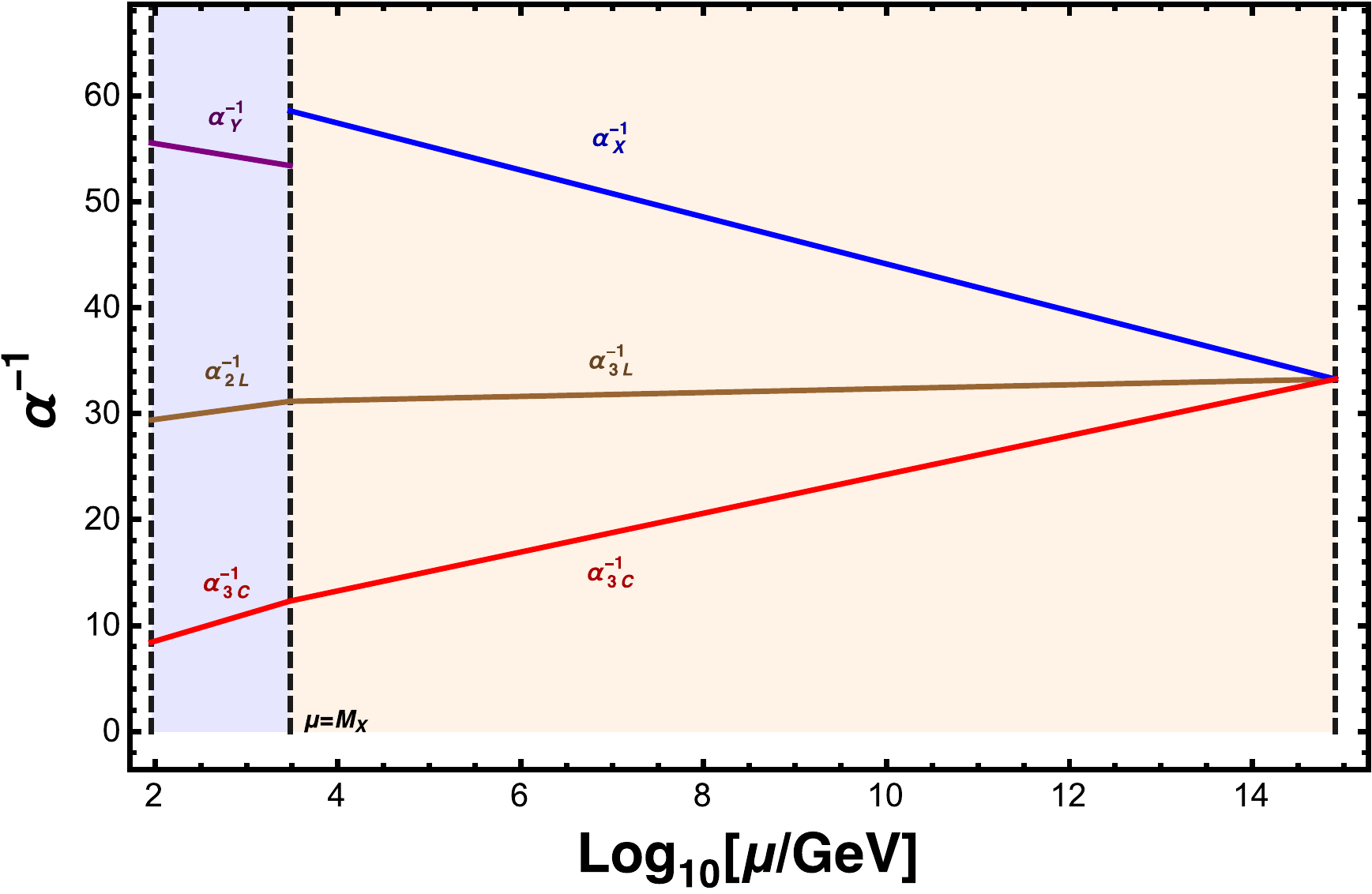}
\caption{Gauge coupling running in the SVS Model adding three
  generations of leptonic octets with \331 symmetry breaking scale at
  $M_{X}= 3000$~GeV, demonstrating successful gauge unification at the
  scale $M_{U} = 10^{14.9}$~GeV with $n_{Y}^{2}=1.8$.}
\label{fig9}
\end{figure}

In Fig.~\ref{fig9} we show an example gauge coupling running with \331
symmetry breaking scale $M_{X}= 3000$~GeV, demonstrating successful
gauge coupling unification at a scale $M_{U} = 10^{14.9}$~GeV with
$n_{Y}^{2}=1.8$. Thus, from the perspective of a low \331 symmetry
breaking scale within the reach of accelerator experiments like the
LHC $\sim{\mathcal{O}}$(TeV)) this model is the most interesting
candidate leading to a successful gauge coupling unification. In
addition to the new gauge bosons, the model can harbor a plethora of
new states associated to the new exotic fermions as well as extra
Higgs bosons.
%


\section{SU(6) Grand Unification}{\label{sec5}}
We consider the possibility of grand unification of the
\331 model in an SU(6) gauge unification group. Two possibilities
arise.  Unlike other conventional grand unified theories, in SU(6) one
can have different components of the 331 subgroup with different
multiplicity. Such a scenario may emerge from the flux breaking of the
unified group in an E(6) F-theory GUT.  This provides new ways of
achieving gauge coupling unification in 331 models.
Alternatively, a sequential variant of \331 model can have a minimal
SU(6) grand unification, which in turn can be a natural E(6)
subgroup. This minimal SU(6) embedding does not require any bulk exotics to
  account for the chiral families and allows for a TeV scale \331
  model.
  
We now demonstrate how the \331 model fermions can be embedded
in an SU(6) grand unified gauge group. Our main consideration is to
explore whether the combinations of the SU(6) gauge group
representations form an anomaly free set, which can contain all the
required fermions. In the subsequent subsections we discuss how
different multiplicities of the SVS version of the \331 model can be
explained when this SU(6) grand unified model is embedded in an E(6)
F-theory and how the sequential \331 model can be embedded in a
minimal anomaly free combination of representations of SU(6) as an
E(6) subgroup. For the minimal SVS version of the \331 model, gauge
coupling unification can be obtained by including both the matter
  multiplets 
in the 27-dimensional fundamental representations of E(6) as
  well as the bulk exotics from the 78-dimensional adjoint
  representations of E(6). In particular the octet of
$\mathrm{SU(3)_L}$ coming from the bulk plays a crucial role in
allowing the unification of the gauge couplings with a low 331
symmetry breaking scale. On the other hand, the embedding of
sequential 331 model in SU(6) does not require any bulk exotics to
account for the chiral multiplets and imply, by adding three
generations of $\mathrm{U(1)_X}$ neutral fermionic octets, one can
obtain SU(6) unification with a TeV scale \331 breaking scale.

We shall first write down some of the product decompositions of the
group SU(6):
\begin{eqnarray}{\label{5.1}}
 6 \times 6 &=& 15_a + 21_s, \nonumber \\
 6 \times \bar{6} &=& 1 + 35, \nonumber \\
 6 \times 15 &=& 20 + 70, \nonumber \\
 6 \times 21 &=& 56 + 70. 
\end{eqnarray}
The SU(6) has \331 as a maximal subgroup with the same rank. For
convenience we write down some of the representations of SU(6) under
this maximal subgroup \331:
\begin{eqnarray}{\label{5.2}}
 6 &=& [3,~1,~-1/3] ~+ ~[ 1,~3,~1/3], \nonumber \\
 15 &=&  [\bar{3},~ 1,~-2/3] ~+~ [1,~\bar{3},~ 2/3] ~+~
 [3,~3,~0],  \nonumber\\
 20 &=& [1,~1,~-1] ~+~ [1,~1,~1] ~+~  [3,~\bar{3},~1/3]~+~ [\bar{3},~3,~-1/3],
 \nonumber\\
 21  &=& [3,~3,~0] ~+~ [6,~ 1,~-2/3] ~+~ [1,~6,~ 2/3],  \nonumber \\
 35  &=& [1,~1,~0] ~+~ [8,~1,~0] ~+~ [1,~8,~0] ~+~ 
 [3,~\bar{3},~-2/3]~+~ [\bar{3},~3,~2/3],
 \nonumber \\
 56  &=&  [10,~1,~-1] ~+~ [1,~10,~1] ~+~ [6,~ 3,~-1/3] ~+~ [3,~6,~ 1/3],
 \nonumber \\
 70  &=&  [6, 3,-1/3] + [3,6, -1/3]  +  [3,\bar{3},1/3]+ 
 [\bar{3},3,-1/3] + [8,1,-1] + [1,8,1]. 
\end{eqnarray}
The anomaly for the various representations of the group SU(6) are
\begin{eqnarray}{\label{5.3}}
 {\cal A}[6] = 1, ~ {\cal A}[15] = 2, ~ {\cal A}[20] = 0, ~
{\cal A}[21] = 10, ~ {\cal A}[35] = 0, ~ {\cal A}[56] = 54, ~
{\cal A}[70] = 27.
\end{eqnarray}
We now turn to two concrete model constructions.

\subsection{SU(6) Grand Unification of the SVS  Model}

It can be easily verified that all fermions of the \331 model proposed
by SVS (discussed in section \ref{sec2}) can be included in the
anomaly free combination of representations under SU(6):
$$ \bar{6} ~+~ \bar{6} ~+~ 15 ~+~ 20.$$
There will be some extra fermions and the multiplicity of the
different representations are now different. It is to be noted that
these states can be naturally embedded in an E(6) theory. We start
with the maximal $\mathrm{SU(2) \times SU(6)}$ subgroup of E(6), and
write down the decomposition:
\begin{eqnarray}{\label{5.4}}
 27 & = & [2,~\bar{6}] ~+~ [1,~ 15] \nonumber \\[.1in]
 78 &=& [1, ~35] ~+~ [2,~20] ~+~ [3,~1] \nonumber
\end{eqnarray}
Thus the \331 anomaly free representations of the SVS model can all be
embedded in a combination of anomaly free representations of SU(6),
which in turn can be embedded in the fundamental and adjoint
representations of the group E(6). The next question is how to match
the multiplicity of the different representations of the SVS 331
model, which is nontrivial. At this stage we resort to the symmetry
breaking at the GUT scale induced by flux breaking through the
Hosotani mechanism~\cite{HOSOTANI1989233}. Assigning particular
geometry to the flux breaking, we identify the different states with
the different algebraic varieties, and then the intersection numbers
would give us the multiplicities of the different representations. A
detailed study of such E(6) F-theory
GUTs~\cite{1126-6708-2009-01-059,King:2010mq,Callaghan:2011jj,Callaghan:2013kaa}
is beyond the scope of this article and we shall rather take a
phenomenological approach to the problem. We consider the required
representations to match the low energy phenomenological
requirements. The first step is to keep the known fermions light and
also to have \331 symmetry breaking scale as low as TeV, while at the
same time requiring for gauge coupling unification.
\begin{figure}[t!]
\includegraphics[width=0.7\linewidth]{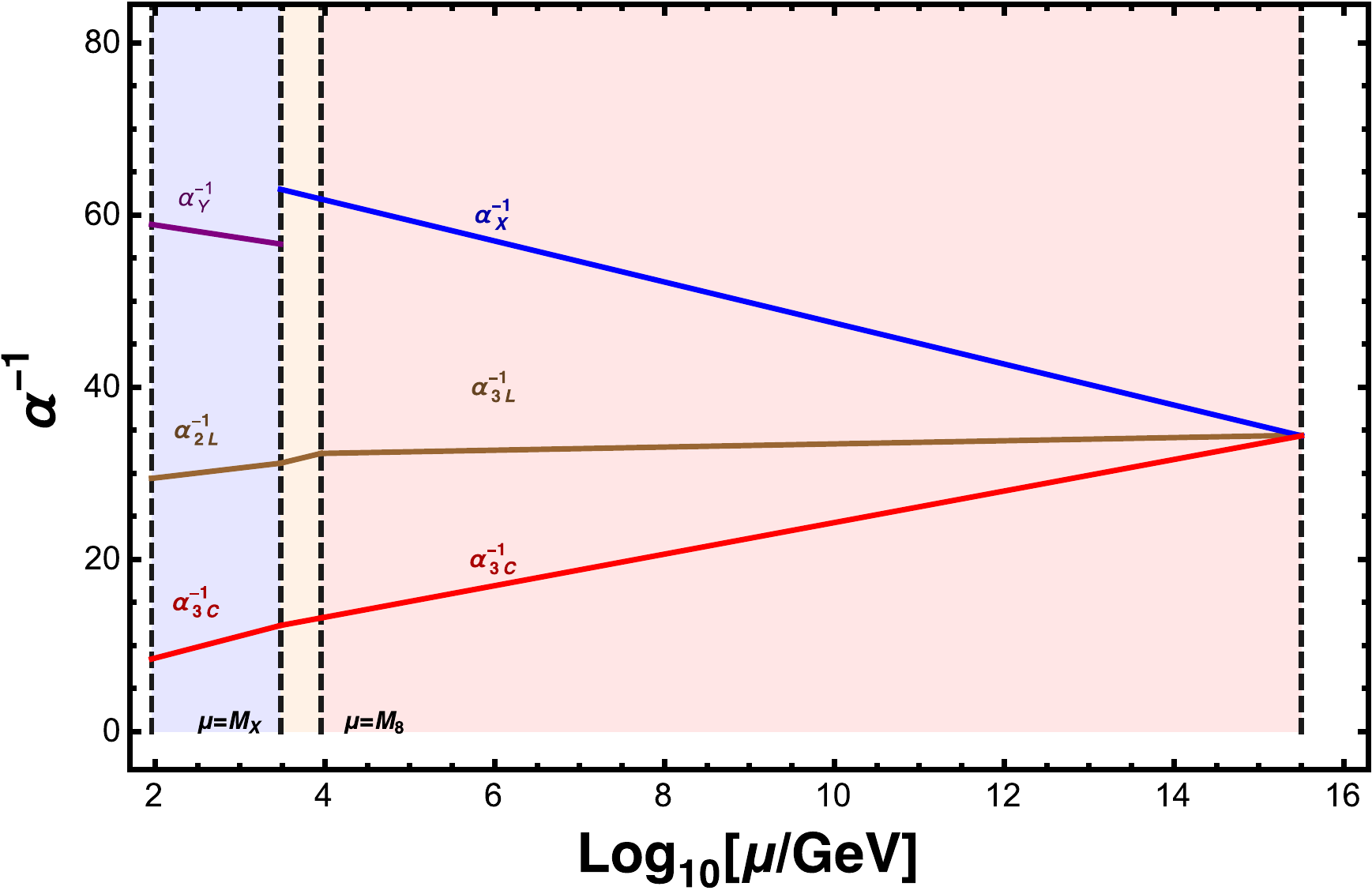}
\caption{Gauge coupling running in the SVS Model with three generations of
  leptonic octets with 331 symmetry breaking scale $M_{X}= 3000$~GeV
  and octet mass scale $M_{8}= 9000$~GeV, demonstrating successful
  gauge unification at the scale $M_{U} = 10^{15.5}$~GeV with
  $n_{Y}=\sqrt{5/3}$ and $n_X=2/\sqrt{3}$.}
\label{svs_unif}
\end{figure}

Considering the $\bar{6}$ representation of SU(6), which contains the
down antiquarks $d^{c}_{L}$ with hypercharge $Y=1/3$, isospin lepton
doublet containing $e_{L}$ and $\nu_{L}$ with $Y=-1/2$, and $N_{L}$
with $Y=0$; we can get the normalization for the hypercharge from
$Tr(Y^2)=5/6 n^{-2}_{Y}$ in the notation of Eq. (\ref{4.7}). The
$\mathrm{U(1)_Y}$ normalization defined in Eq. (\ref{4.7}) is given by
$n_Y=\sqrt{5/3}$ and using Eq. (\ref{4.8}) we obtain the
$\mathrm{U(1)_X}$ normalization given by $n_X=2/\sqrt{3}$, which is
below the normalizations required for a guaranteed unification in
Fig. \ref{fig1} and Fig. \ref{fig7}. However, it is still possible to
obtain gauge coupling unification following the prescription in
Ref. \cite{Boucenna:2014dia}, where the octet scale is decoupled from
the \331 symmetry breaking scale and is assumed to lie between the
\331 symmetry breaking scale and unification scale. Here, the octets
belong to 35 of SU(6), which belongs to the bulk exotics coming from
the 78-dimensional adjoint representations of E(6).

Using Eq.~(\ref{4.3}) the the one-loop beta-coefficients $b_i$ can be
calculated for the different phases. For the phase between the
electroweak symmetry breaking scale and the \331 symmetry breaking
scale ($M_Z$ to $M_X$) the one-loop beta-coefficients are given by
$b_{2L} = -19/6$, $b_{Y} =41/10$, $b_{3C} = -7$. For the phase between
the \331 symmetry breaking scale and the octet mass scale ($M_X$ to
$M_8$) the one-loop beta-coefficients are given by $b_{3L} = -13/2$,
$b_{X} =13/2$, $b^{331}_{3C} = -5$. Finally, for the phase between the
octet mass scale to the unification scale ($M_8$ to $M_U$) the
one-loop beta-coefficients are given by $b^{8}_{3L} =2n-13/2 $, where
$n$ is the number of generations of the fermionic octets
($ \Omega\equiv [1, 8^{\ast},0]$), $b^{8}_{X} =13/2$,
$b^{8}_{3C} = -5$.  

In Fig.~\ref{svs_unif} we plot the gauge coupling
running of SVS \331 model with the field content given in
Eqs.~(\ref{2.3},\ref{2.4}) and three generations of fermionic octets with \331 symmetry breaking scale
$M_{X}= 3000$~GeV and octet mass scale $M_{8}= 9000$~GeV,
demonstrating successful gauge unification at the scale
$M_{U} = 10^{15.5}$~GeV with $n_{Y}=\sqrt{5/3}$ and $n_X=2/\sqrt{3}$. A relative modest variation of the octet mass scale from the \331 scale, $M_8 / M_X = 3$, therefore lifts the scale of successful unification from the value $M_{U} = 10^{14.9}$~GeV found in Fig.~\ref{fig9} and thus relaxes the tension with proton decay limits, cf. Section~\ref{sec6}. 


\subsection{SU(6) Grand Unification of the sequential \331 Model}

It is easy to verify from Eq. (\ref{5.1}) that each generation of the
fermionic multiplets of the sequential 331 model written in
Eq. (\ref{3.1}) fits perfectly in the anomaly free combination of
SU(6) representations: $ \bar{6} ~+~ \bar{6^{\prime}} ~+~ 15 $, where
$\bar{6}$ contains $d^{c}_{L}\equiv [3,1,-1/3]$ and
$\psi_{L}\equiv [1,3^{\ast},-1/3]$; $\bar{6^{\prime}}$ contains
$D^{c}_{L}\equiv [3,1,-1/3]$ and $\xi_{L}\equiv [1,3^{\ast},-1/3]$;
and $15$ contains $u^{c}_{L}\equiv [3^{\ast},1,-2/3]$,
$\chi_{L}\equiv [1,3^{\ast},2/3]$ and $Q_{L}\equiv [3,3,0]$. Now the
fundamental $27$ of E(6) branches under the maximal
$\mathrm{SU(2) \otimes SU(6)}$ subgroup as
$27 = [2,~\bar{6}] ~+~ [1,~ 15]$. Thus three $27$s of E(6) contain
three sets of $ \bar{6} ~+~ \bar{6^{\prime}} ~+~ 15 $ accommodating
the three generations of the fermionic multiplets of the sequential
331 model. However the minimal content of the sequential 331 model
does not have a low scale unification. However, by adding three
generations of fermionic octets
again leads to a successful gauge coupling unification. 
\begin{figure}[t!]
\includegraphics[width=0.7\linewidth]{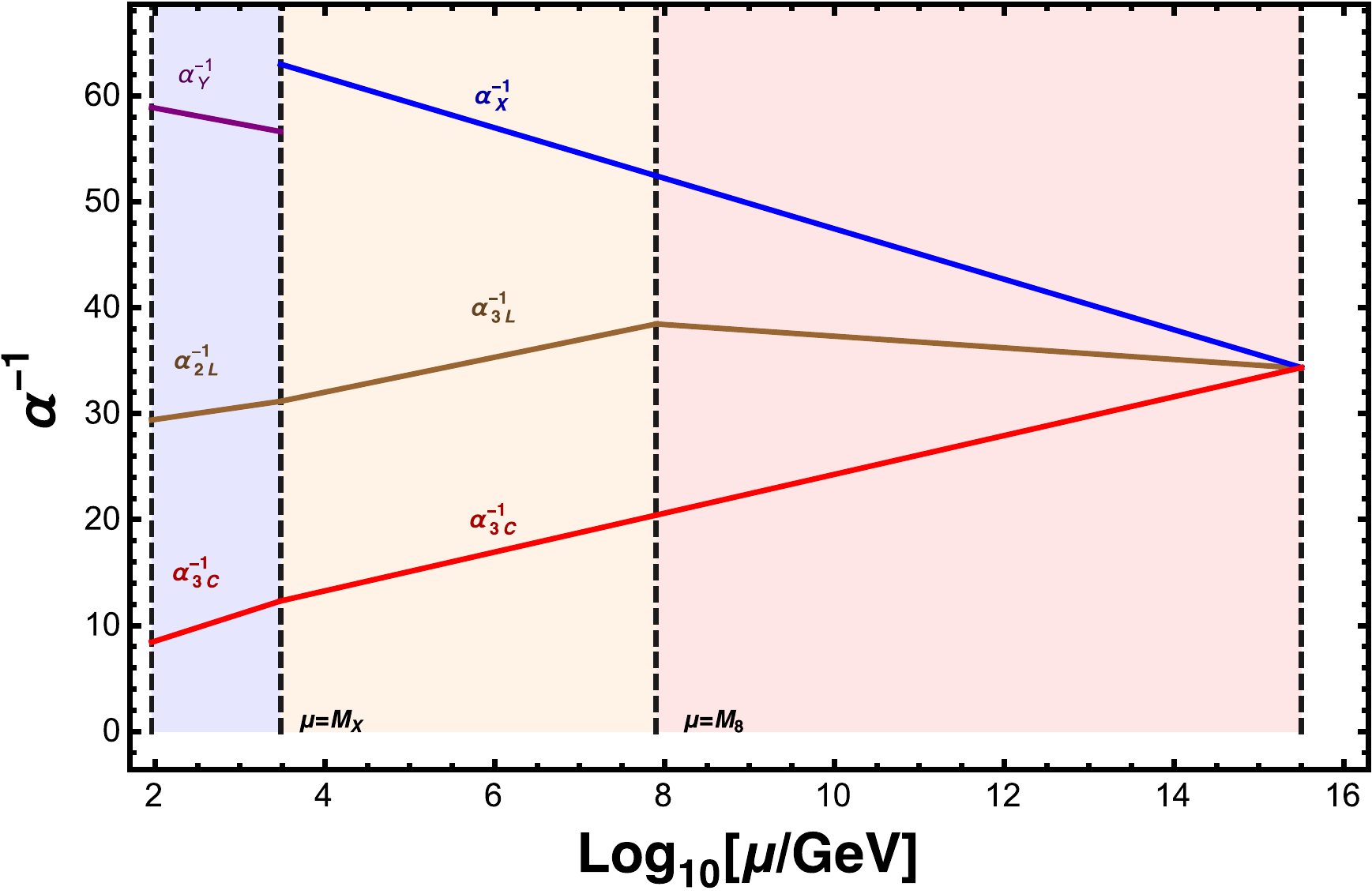}
\caption{Gauge coupling running in the sequential 331 Model with three
  generations of fermionic octets with \331 symmetry breaking scale
  $M_{X}= 3000$~GeV and octet mass scale $M_{8}= 8\times 10^7$~GeV,
  demonstrating successful gauge unification at the scale $M_{U} =
  10^{15.5}$~GeV with $n_{Y}=\sqrt{5/3}$ and $n_X=2/\sqrt{3}$.}
\label{seq_unif}
\end{figure}

For the
phase between the electroweak symmetry breaking scale and the \331
symmetry breaking scale ($M_Z$ to $M_X$) the one-loop
beta-coefficients are given by $b_{2L} = -19/6$, $b_{Y} =41/10$,
$b_{3C} = -7$. For the phase between the \331 symmetry breaking scale
and the octet mass scale ($M_X$ to $M_8$) the one-loop
beta-coefficients are given by $b_{3L} = -9/2$, $b_{X} =13/2$,
$b^{331}_{3C} = -5$. Finally, for the phase between the octet mass
scale to the unification scale ($M_8$ to $M_U$) the one-loop
beta-coefficients are given by $b^{8}_{3L} =2n-9/2 $, where $n$ is the
number of generations of the fermionic octets
($ \Omega\equiv [1, 8^{\ast},0]$), $b^{8}_{X} =13/2$,
$b^{8}_{3C} = -5$. 

In Fig. \ref{seq_unif} we plot the gauge coupling
running of the sequential \331 model with the field content given in
Eqs. (\ref{2.4},\ref{3.1}) and three generations of fermionic octets with \331 symmetry breaking scale
$M_{X}= 3000$~GeV and octet mass scale $M_{8}= 8\times 10^7$~GeV,
demonstrating successful gauge unification at the scale
$M_{U} = 10^{15.5}$~GeV with $n_{Y}=\sqrt{5/3}$ and $n_X=2/\sqrt{3}$. In this scenario, the octet mass scale has to be detached rather strongly from the \331 scale in order to achieve successful unification.


\section{ $\sin^{2}\theta_{w}$ and proton decay  in SU(6) grand unification}
\label{sec6}

Using the RGEs and the relations among the coupling constants
corresponding to different gauge groups one can express
$\sin^{2}\theta_{w} (M_Z)$ in terms of the different scales associated
with the SU(6) grand unified theory. Noting that for SU(6) grand
unification we have $n_{Y}=\sqrt{5/3}$ and $n_X=2/\sqrt{3}$, the
relation between normalized couplings at the scales $M_Z$ and $M_X$
are given by
\begin{eqnarray}{\label{sin:1.1}}
\alpha_{2L}^{-1}(M_Z) &=& \alpha_{\rm{em}}^{-1}(M_Z)-\frac{5}{3} {\alpha^{N}_{Y}}^{-1}(M_Z), \\
{\alpha^{N}_{Y} }^{-1} (M_X) &=& \frac{1}{5}\alpha_{3L}^{-1} (M_X) +\frac{4}{5}{\alpha^{N}_{X}}^{-1} (M_X).
{\label{sin:1.2}}
\end{eqnarray}
Using Eq.~(\ref{sin:1.1}) it is straightforward to obtain
\begin{equation}{\label{sin:1.3}}
\sin^{2} \theta_w (M_Z) \equiv \frac{\alpha_{\rm{em}}(M_Z)}{\alpha_{2L}(M_Z)}=\frac{3}{8}+\frac{5}{8} \alpha_{\rm{em}}(M_Z) \left[ \alpha_{2L}^{-1} (M_Z) - {\alpha^{N}_{Y} }^{-1} (M_Z)\right] .
\end{equation}
Finally, using Eqs. (\ref{4.2}), (\ref{sin:1.2}) the above equation can be written in the form
\begin{eqnarray}{\label{sin:1.3}}
\sin^{2} \theta_w (M_Z) = &&\frac{3}{8}+\frac{5}{8} \alpha_{\rm{em}}(M_Z) \left[ \frac{4}{5} \left\{ \frac{b_{3L}}{2\pi} \ln \left( \frac{M_8}{M_X}\right) +\frac{b_{3L}^8}{2\pi} \ln \left( \frac{M_U}{M_8}\right) \right\} \right.\nonumber\\
&& \left. + \frac{(b_{2L}-b_Y)}{2\pi} \ln \left( \frac{M_X}{M_Z}\right) -\frac{4}{5} \frac{b_{X}}{2\pi} \ln \left( \frac{M_U}{M_X}\right)  \right] ,
\end{eqnarray}
which can be readily used to obtain the prediction for
$\sin^{2}\theta_{w} (M_Z)$. For example, in the sequential 331 model
taking \331 symmetry breaking scale $M_{X}= 3000$~GeV, octet mass
scale $M_{8}= 8\times 10^7$~GeV, and unification scale
$M_{U} =10^{15.5}$~GeV we obtain
$\sin^{2}\theta_{w} (M_Z) \simeq 0.231$, which is consistent with the
electroweak precision data \cite{Agashe:2014kda}.

Turning to the prediction for proton decay, we note that being a
non-supersymmetric scenario the gauge $d=6$ contributions for proton
decay are most important here. An analysis of all \SM invariant
operators
\cite{Weinberg:1979sa,Weinberg:1980bf,Wilczek:1979hc,Weldon:1980gi}
that can induce proton decay in SU(6) is beyond the scope of this
article and will be addressed in a separate communication. Here we
will consider the decay mode $p\rightarrow e^{+}\pi^{0}$, which is
constrained by experimental searches to have a life time
$\tau^{\rm{expt}}_{p}\geq 1 \times 10^{34}$ \cite{Agashe:2014kda}. The
relevant effective operators in the physical basis are given by
\cite{FileviezPerez:2004hn,Nath:2006ut}
\begin{eqnarray}{\label{pd:6.1}}
\mathcal{O}(e^{c}_{\alpha},d_{\beta}) &=& c (e^{c}_{\alpha},d_{\beta}) \epsilon_{ijk} \overline{u^{c}_{i}}_{L} \gamma^{\mu} u_{jL} \overline{e^{c}_{\alpha}}_{L} \gamma_{\mu} {d_{k\beta}}_{L} ,
 \nonumber\\
 \mathcal{O}(e_{\alpha},d^{c}_{\beta}) &=& c (e_{\alpha},d^{c}_{\beta}) \epsilon_{ijk} \overline{u^{c}_{i}}_{L} \gamma^{\mu} u_{jL} \overline{d^{c}_{k\beta}}_{L} \gamma_{\mu} {e_{\alpha}}_{L} ;
\end{eqnarray}
where 
\begin{eqnarray}{\label{pd:6.2}}
 c (e^{c}_{\alpha},d_{\beta})  &=& k_1^2 \left[ V^{11}_{1} V^{\alpha \beta}_{2} +(V_1 V_{UD})^{1\beta} (V_2 V_{UD}^{\dagger})^{\alpha 1}  \right] ,
 \nonumber\\
c (e_{\alpha},d^{c}_{\beta})  &=&  k_1^2 V^{11}_{1} V^{ \beta \alpha}_{3}+k^2_2 (V_4 V_{UD}^{\dagger})^{\beta 1}+(V_1 V_{UD} V_4^{\dagger} V_3)^{1\alpha} ;
 \nonumber\\
 \alpha &=& \beta \neq 2.
\end{eqnarray}
Here $i,j,k=1,2,3$ are the color indices and $\alpha,\beta =1,2$;
$V_{1,2,3,4}$ and $V_{UD}$ are the mixing matrices
$V_1=U^{\dagger}_{C} U$, $V_2=E^{\dagger}_{C}D$,
$V_3=D^{\dagger}_{C}E$, $V_{UD}=U^{\dagger}D$; where $U,D,E$ are the
unitary matrices diagonalizing the Yukawa couplings e,g.
$U^{T}_{C}Y_U U=Y_{U}^{\rm{diag}}$. $k_1=g_{\rm{GUT}}/M_{(X,Y)}$ and
$k_2=g_{\rm{GUT}}/M_{(X^{\prime},Y^{\prime})}$, where
$M_{(X,Y)}, M_{(X^{\prime},Y^{\prime})}\sim M_{\rm{GUT}}$ are the
masses of the superheavy gauge bosons and $g_{\rm{GUT}}$ is the
coupling constant at the GUT scale. The decay rate for
$p\rightarrow e^{+}\pi^{0}$ mode is given by
\begin{eqnarray}{\label{pd:6.3}}
\Gamma (p\rightarrow e^{+}\pi^{0})=\frac{m_p}{16\pi f_{\pi}^{2}}A_{L}^{2} \vert \alpha_H \vert^{2} (1+D+F)^{2} \left[  \vert c (e,d^{c})\vert^{2}  +   \vert  c (e^{c},d) \vert^{2}  \right],
\end{eqnarray}
where $m_p=938.3$ MeV is the proton mass, $f_{\pi}=139$ MeV is the
pion decay constant, $A_{L}$ is the long distance renormalization
factor; $D,F$ and $\alpha_H$ are parameters of the chiral
Lagrangian. For a rough estimate, taking
$\alpha_H(1+D+F)\sim 0.012 ~{\rm{GeV}}^3$
\cite{Aoki:2006ib,Aoki:2008ku}; $A_R=A_L A^{\rm{SD}}_R \sim 3$, where
$A^{\rm{SD}}_R$ is the short distance renormalization factor; the
parameter depending on the mixing matrices $F_{q}(V)\sim 5$, we obtain

\begin{eqnarray}{\label{pd:6.4}}
\Gamma^{-1} (p\rightarrow e^{+}\pi^{0}) \sim 10^{36}\text{ yrs}
\left(\frac{\alpha^{-1}_\text{GUT}}{35}\right)^2
\left(\frac{M_U}{10^{16}\text{ GeV}}\right)^4.
\end{eqnarray}

Now noting that $M_{U} =10^{15.5}$~GeV and
$\alpha^{-1}_{\rm{GUT}}\sim35$ in the SVS and sequential 331 models,
the lifetime of the proton decay mode $p\rightarrow e^{+}\pi^{0}$
comes out to be \footnote{In fact for a more careful estimation, one
  should also take into account the GUT threshold corrections which
  might improve on this limit, however, given the uncertainties in the
  hadronic parameters here we do not worry about such effects.}
$\sim 10^{34}$~yrs, which is consistent with the current
experimental limit \cite{Agashe:2014kda}.

\section{Discussion and outlook}
\label{sec:discussion-outlook}

In this paper we have considered the possibility of conventional
non-supersymmetric grand unification of extended electroweak models
based upon the \331 gauge framework within an SU(6) gauge unification
group.
In contrast to other conventional grand unified theories, in SU(6) one
can have different components of the \331 subgroup with different
multiplicity.
Such scenarios may emerge from the flux breaking of the unified group
in an E(6) F-theory GUT framework. While it allows for successful
unification, the required 331 scale is typically very close to
unification. 

However, the sequential addition of a leptonic octet provides a way of
achieving gauge coupling unification at 331 scales accessible at
collider experiments.
Alternatively, we have also considered a sequential variant of the
\331 model that can have a minimal SU(6) grand unification, which in
turn can be a natural E(6) subgroup.
Such minimal SU(6) embedding does not require any bulk exotics in
order to account for the chiral families and allows for a TeV scale
\331 model as well as seesaw-induced neutrino masses.

In both cases the gauge coupling unification is associated to
the presence of sequential a leptonic octet at some intermediate
scale between the 331 scale, which lies in the TeV range, and the
unification scale.
It is important to stress that the presence of the octet plays a key
role in the mechanism of neutrino mass generation. In other words,
the same physics that drives unification is responsible for the
radiative origin of neutrino masses~\cite{Boucenna:2014dia}.

\section*{Acknowledgements}

This work is supported by the Spanish grants FPA2014-58183-P,
Multidark CSD2009-00064, SEV-2014-0398 (MINECO) and
PROMETEOII/2014/084 (GVA). CH would like to thank the organizers of
Planck 2016, Valencia, for their warm hospitality and IFIC's AHEP
Group, Institut de Fisica Corpuscular -- C.S.I.C./Universitat de
Valencia, Spain, where part of this work was carried out. The work of SP is partly supported by DST, India under the financial grant SB/S2/HEP-011/2013. The work of US is supported partly by the JC Bose National Fellowship grant under DST, India.

\bibliography{merged_Valle,newrefs,z4,protondecay}

\end{document}